\documentclass[aps,prd,preprint,superscriptaddress,tightenlines,nofootinbib]{revtex4}
\usepackage{epsfig}
\usepackage{graphicx}
\usepackage{dcolumn}
\usepackage{bm}

\newcommand{\etal}{{\it et al.}}

\begin{document}

\preprint{\tighten\vbox{\hbox{\hfil CLNS 06/1958}
                        \hbox{\hfil CLEO 06-07}}}

\title{\LARGE An Investigation of $D^+\to\tau^+\nu$}

\author{P.~Rubin}
\affiliation{George Mason University, Fairfax, Virginia 22030}
\author{C.~Cawlfield}
\author{B.~I.~Eisenstein}
\author{I.~Karliner}
\author{D.~Kim}
\author{N.~Lowrey}
\author{P.~Naik}
\author{C.~Sedlack}
\author{M.~Selen}
\author{E.~J.~White}
\author{J.~Wiss}
\affiliation{University of Illinois, Urbana-Champaign, Illinois
61801}
% ...
\collaboration{CLEO Collaboration} 
\noaffiliation

\date{April 21, 2006)}

\begin{abstract}

We test whether or not the $\tau$ lepton manifests the same
couplings as the $\mu$ lepton by investigating the relative decay
rates in purely leptonic $D^+$ meson decays. Specifically, we
place the first upper limit on the ratio
 $R=\Gamma\left(D^+\to\tau^+\nu\right)/\Gamma\left(
D^+\to\mu^+\nu\right)$. We use 281 pb$^{-1}$ of data accumulated at
the $\psi(3770)$ resonance with the CLEO-c detector, to determine
${\cal{B}}(D^+\to\tau^+\nu)<2.1\times 10^{-3}$ at 90\% confidence
level (C. L.).  The ratio of $R$ to the Standard Model expectation
of 2.65 then is $<$1.8 at 90\% C. L., consistent with the prediction
of lepton universality.
\end{abstract}


\maketitle \tighten

\section{Introduction}

\begin{figure}[htbp]
 \vskip 0.00cm
 \centerline{ \epsfxsize=3.0in \epsffile{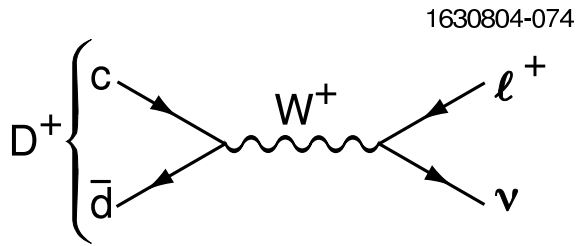} }
 \caption{The decay diagram for $D^+\to \ell^+\nu$.} \label{Dptomunu}
 \end{figure}

The Standard Model decay diagram for $D^+\to \ell^+\nu$ is shown
in Fig.~\ref{Dptomunu}. The decay rate is given by \cite{Formula1}
\begin{equation}
\Gamma(D^+\to \ell^+\nu) = {{G_F^2}\over
8\pi}f_{D^+}^2m_{\ell}^2M_{D^+} \left(1-{m_{\ell}^2\over
M_{D^+}^2}\right)^2 \left|V_{cd}\right|^2~~~, \label{eq:equ_rate}
\end{equation}
where $M_{D^+}$ is the $D^+$ mass, $m_{\ell}$ is the mass of the
charged final state lepton, $V_{cd}$ is a
Cabibbo-Kobayashi-Maskawa matrix element with a value we take
equal to 0.225 \cite{KTeV}, and $G_F$ is the Fermi coupling
constant.

The decay is helicity suppressed because the virtual $W^+$ is a
spin-1 particle, and the final state consists of a naturally
left-handed spin-1/2 neutrino and a naturally right-handed spin-1/2
anti-lepton that have equal energies and opposite momenta. The ratio
of decay rates for any two different leptons is then fixed by
well-known masses. For example, for $\tau^+\nu$ to $\mu^+\nu$, the
expected ratio is

\begin{equation}
R\equiv \frac{\Gamma(D^+\to \tau^+\nu)}{\Gamma(D^+\to \mu^+\nu)}=
{{m_{\tau^+}^2 \left(1-{m_{\tau^+}^2\over
M_{D^+}^2}\right)^2}\over{m_{\mu^+}^2 \left(1-{m_{\mu^+}^2\over
M_{D^+}^2}\right)^2}}~~. \label{eq:rat}
\end{equation}

Any deviation from this formula would be a manifestation of physics
beyond the Standard Model. This could occur if any other charged
intermediate boson existed that coupled to leptons differently than
mass-squared. Then the couplings would be different for muons and
$\tau$'s. This would be a manifest violation of lepton universality,
which has identical couplings of the muon, the tau, and the electron
to the gauge bosons ($\gamma,~Z^0$ and $W^{\pm}$) \cite{Hewett}. (We
note that in some models of supersymmetry the charged Higgs boson
couples as mass-squared to the leptons and therefore its presence
would not cause a deviation from Eq.~\ref{eq:rat} \cite{Hou}.) Using
measured masses \cite{PDG}, this expression yields a value of 2.65
with a negligibly small error.

We have already reported \cite{munuPRL}
${\cal{B}}(D^+\to\mu^+\nu)=(4.40\pm 0.66^{+0.09}_{-0.12})\times
10^{-4},$ and established an upper limit of ${\cal{B}}(D^+\to
e^+\nu)<2.4\times 10^{-5}$. It remains to measure or limit
$\tau^+\nu$, which is the subject of this paper. We note, for
reference, that the predicted relative widths in the Standard
Model are $2.65:1:2.3\times 10^{-5}$ for the $\tau^+ \nu$, $\mu^+
\nu$ and $e^+ \nu$ final states, respectively.

The CLEO-c detector is equipped to measure the momenta and
directions of charged particles, identify charged hadrons, detect
photons and determine with good precision their directions and
energies. It has been described in more detail previously
\cite{CLEODR}. Particle identification is accomplished using both
dE/dx information in the tracking drift chamber and in a separate
Ring Imaging Cherenkov Detector (RICH) \cite{fakes}.

\section{Data Sample and Signal Selection Overview}

In this study we use 281 pb$^{-1}$ of CLEO-c data produced in
$e^+e^-$ collisions and recorded at the $\psi(3770)$ resonance. At
this energy, the event sample consists of a mixture of pure
$D^+D^-$, $D^0\overline{D}^0$, three-flavor continuum, and
$\gamma\psi(2S)$ events. There are also $\tau^+\tau^-$ pairs,
two-photon events, and non-$D\overline{D}$ decays of the
$\psi(3770)$, whose production rates are small enough for them not
to contribute background in this study.


This analysis follows very closely our previous study of
$D^+\to\mu^+\nu$ \cite{munuPRL,CLEODptomunu}. First we fully
reconstruct a sample of hadronic $D^-$ decays, that we call tags,
and then search for tracks that are consistent with a $\pi^+$ from
the decay sequence $D^+\to\tau^+\nu$,
$\tau^+\to\pi^+\overline{\nu}$, rather than a muon directly from
two-body $D$ decay.  Besides using $D^-$ tags and searching for
$D^+\to\tau^+\nu$, $\tau^+\to\pi^+\overline{\nu}$ we also use the
charge-conjugate $D^+$ tags and search for $D^-\to
\tau^-\overline{\nu}$, $\tau^-\to\pi^-{\nu}$; in the rest of this
paper we will not mention the charge-conjugate modes explicitly,
but they are always used. The loss of rate compared to the muon
case, caused by the ${\cal{B}}(\tau^+\to\pi^+\nu)$ of
(11.06$\pm$0.11)\% \cite{PDG}, is somewhat compensated for by the
larger $D^+\to\tau^+\nu$ branching ratio as given by
Eq.~\ref{eq:equ_rate}. This search has a smeared signal region as
compared to the muon case because of the extra missing neutrino,
and therefore backgrounds are a much more serious concern.

We examine all the recorded events and retain those containing at
least one charged $D$ candidate in the modes listed in
Table~\ref{tab:Dreconnew}. Track selection, particle
identification, $\pi^0$, $K_S$ and muon selection cuts are
identical to those described in Ref. \cite{CLEODptomunu}.

We have investigated using other $\tau$ decay modes but they all
have significant problems. The semileptonic mode $e\nu\bar{\nu}$ is
embedded in a large $D^+$ semileptonic background. The
$\rho^+\overline{\nu}$ mode has a MM$^2$ resolution approximately
twice as poor, and the $\pi^+\pi^+\pi^-\overline{\nu}$ mode has
several additional associated backgrounds, for example
$D^+\to\pi^+\pi^+\pi^-\pi^0$ and $\eta\mu^+\nu$, that severely limit
its usefulness.

\section{Reconstruction of Charged ${\boldmath D}$ Tagging Modes}

Tagging modes are fully reconstructed by first evaluating the
difference in the energy, $\Delta E$, of the decay products with the
beam energy.  We require the absolute value of this difference to
contain 98.8\% of the signal events, i.e. to be within approximately
2.5 times the r.m.s width of the peak value. The r.m.s. widths vary
from 7 MeV in the $K^+K^-\pi^-$ mode to 14 MeV in the
$K^+\pi^-\pi^-\pi^0$ mode. For the selected events we then calculate
the reconstructed $D^-$ beam-constrained mass defined as
\begin{equation}
m_{\mathrm{BC}}=\sqrt{E_{\mathrm{beam}}^2-(\sum_i\overrightarrow{p}_{\!i})^2},
\end{equation}
where $i$ runs over all the final state particles. The
beam-constrained mass has better resolution than merely
calculating the invariant mass of the decay products since the
beam has a small energy spread.

\begin{figure}[htbp]
\centerline{ \epsfxsize=6.0in \epsffile{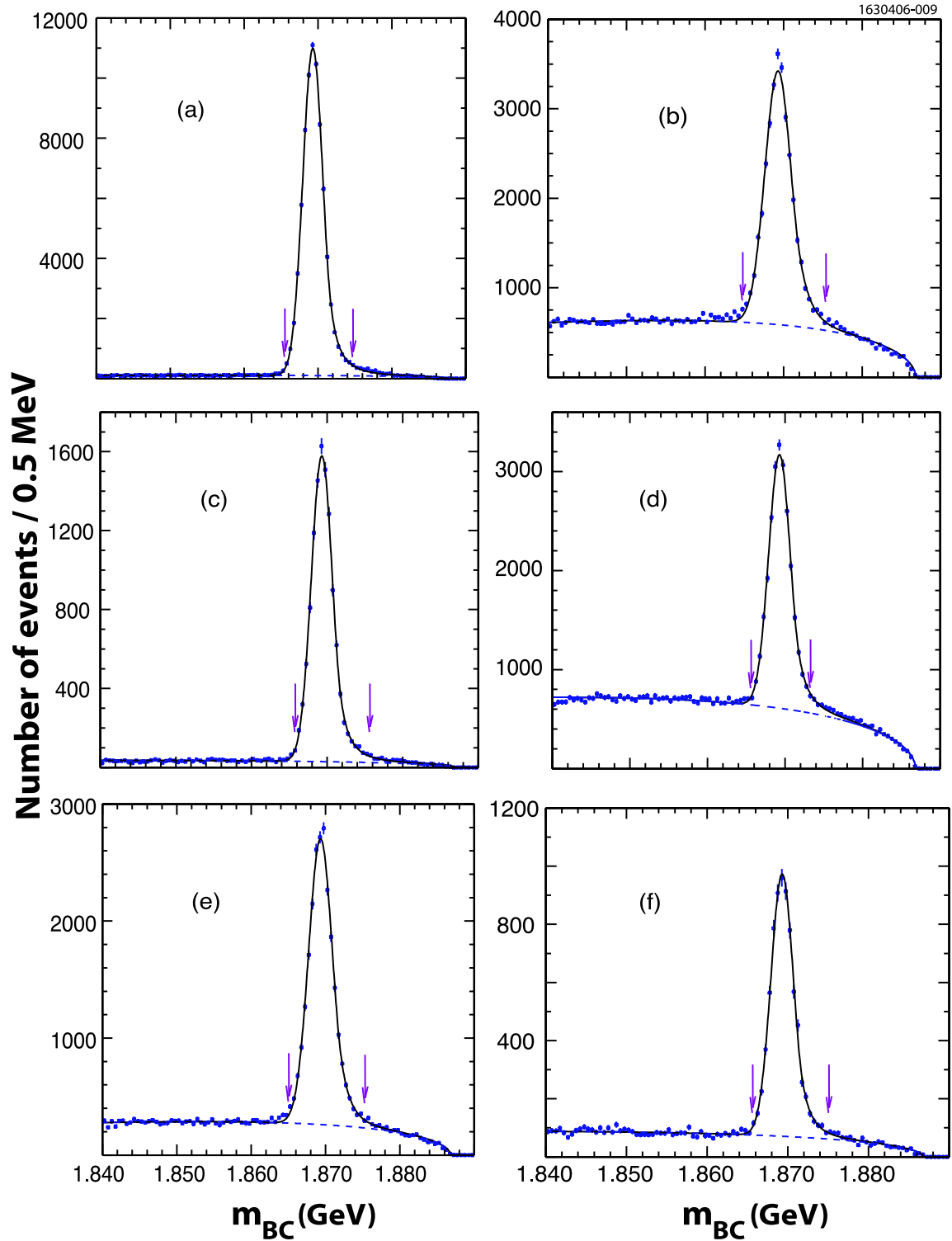} }
\caption{ Beam-constrained mass distributions for different fully
reconstructed $D^-$ decay candidates in the final states: (a) $K^+
\pi^- \pi^-$, (b) $K^+ \pi^- \pi^- \pi^0$, (c) $K_S\pi^-$, (d)
$K_S \pi^-\pi^-\pi^+$, (e) $K_S\pi^- \pi^0$ and (f) $K^+K^-\pi^-$.
The solid curves show the sum of signal and background functions.
The dashed curves indicate the background fits. Events between the
arrows are selected for further analysis.} \label{Dreconnew}
\end{figure}

The $m_{\mathrm{BC}}$ distributions for all $D^-$ tagging modes
considered in this data sample are shown in Fig.~\ref{Dreconnew}.
They are listed in Table~\ref{tab:Dreconnew}, along with the
numbers of signal events and background events within the regions
shown by the arrows in Fig.~\ref{Dreconnew}. The tag candidates
are subjected to $\Delta E$ and $m_{\mathrm{BC}}$ cuts explained
in our previous paper \cite{munuPRL}. The numbers of tagged events
are determined from fits of the $m_{\mathrm{BC}}$ distributions to
a signal function plus a background shape. For the background we
fit with a shape function analogous to one first used by the ARGUS
collaboration \cite{ARGUS}, that has approximately the correct
threshold behavior at large $m_{\mathrm{BC}}$. To use this
function, we first fit it to the data selected by using $\Delta E$
sidebands, defined as $5\sigma<|\Delta E|< 7.5\sigma$, where
$\sigma$ is the r.m.s. width of the $\Delta E$ distribution. Doing
this mode by mode and allowing the normalization to float, we fix
the shape parameters. For the signal we use a lineshape similar to
that used for extracting photon signals from electromagnetic
calorimeters because of the tail towards high mass caused by
initial-state radiation \cite{CBL}. The functional form is
\[f(m_{\mathrm{BC}}|m_D,\sigma_{m_{\mathrm{BC}}},\alpha,n)= \left( \begin{array}{l}
  A\cdot{\rm exp}\left[-{1\over 2}\left({{m_{\mathrm{BC}}-m_D}\over \sigma_{m_{\mathrm{BC}}}}
\right)^2\right]~~~~{{\rm for}~m_{\mathrm{BC}}<m_D-\alpha\cdot\sigma_{m_{\mathrm{BC}}}}\\
 A\cdot{{\left({n\over \alpha}\right)^n e^{-{1\over 2}\alpha^2}
\over \left({{m_{\mathrm{BC}}-m_D}\over
\sigma_{m_{\mathrm{BC}}}}+{n\over \alpha}-\alpha\right)^n}}
~~~~~~~~~~~{{\rm for}~m_{\mathrm{BC}}>m_D-\alpha\cdot\sigma_{m_{\mathrm{BC}}}}\\
{\rm here}~A^{-1}\equiv \sigma_{m_{\mathrm{BC}}}\cdot
\left[{n\over \alpha}\cdot {1\over {n-1}}e^{-{1\over 2}\alpha^2}
+\sqrt{\pi\over 2}\left(1+{\rm
erf}\left({\alpha\over\sqrt{2}}\right) \right)\right]
\end{array}\right.\]
\begin{equation}
\end{equation}
Here $m_{\mathrm{BC}}$ is the measured mass of each candidate, $m_D$
is the ``true'' (or most likely) mass, $\sigma_{m_{\mathrm{BC}}}$ is
the mass resolution, and $n$ and $\alpha$ are parameters governing
the shape of the high mass tail. All these quantities are allowed to
float in the separate fits of each mode.

\begin{table}[htb]
\begin{center}
\begin{tabular}{lcc}
    Mode  &  Signal           &  Background \\ \hline
$K^+\pi^-\pi^- $ & $77387 \pm 281$   & $~1868$\\
$K^+\pi^-\pi^- \pi^0$ & $24850 \pm 214$  & $12825$\\
$K_S\pi^-$ &   $11162\pm 136$& $~~514$\\
$K_S\pi^-\pi^-\pi^+ $ &  $18176 \pm 255$ & $~8976$\\
$K_S\pi^-\pi^0 $ &  $20244\pm 170$ & $5223$\\
$K^+K^-\pi^-$ & 6535$\pm$~95 &~1271 \\\hline
Sum &  $ 158354\pm 496$ & $30677$\\
\hline\hline
\end{tabular}
\end{center}
\caption{Tagging modes and numbers of signal and background events
determined from the fits shown in Fig.~\ref{Dreconnew}.
\label{tab:Dreconnew}}
\end{table}

We use a total of 158,354$\pm$496$\pm$475 single tag events for
further analysis. The systematic error on this number is
determined by varying the background function and is estimated at
0.5\%.

\section{${\boldmath D^+\to \tau^+\nu}$ Selection Criteria}
\label{sec:muonsel}

As in our search for $D^+\to\mu^+\nu$, we calculate the missing
mass squared (MM$^2$) defined as
\begin{equation}
{\rm
MM}^2=\left(E_{\mathrm{beam}}-E_{\mathrm{track}}\right)^2-\left(-\overrightarrow{p}_{\!D^-}
-\overrightarrow{p}_{\!\mathrm{track}}\right)^2, \label{eq:MMsq}
\end{equation}
where $E_{\mathrm beam}$ is the beam energy, $E_{\mathrm track}$ and
$\left(-\overrightarrow{p}_{\!D^-}
-\overrightarrow{p}_{\!\mathrm{track}}\right)$ are the measured
energy and momentum of a single track, assuming that the track is a
pion, and $\overrightarrow{p}\!_{D^-}$ is the three-momentum of the
fully reconstructed $D^-$.

\begin{figure}[htb]
 \centerline{
 \epsfxsize=4.0in \epsffile{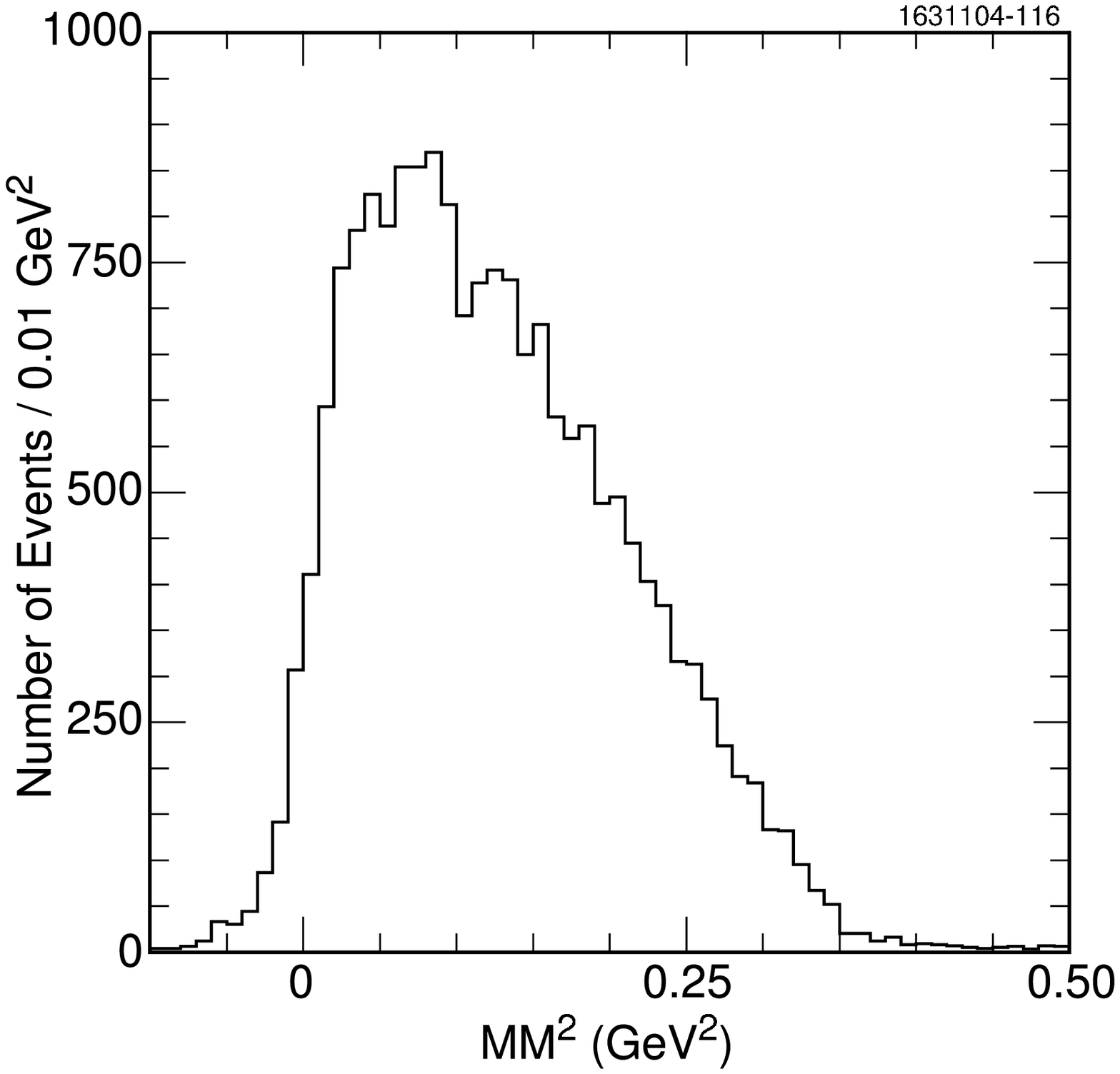}}
 \caption{Missing
 mass squared distribution for $D^+ \to \tau^+\nu$, $\tau^+\to
 \pi^+\overline{\nu}$ from Monte Carlo simulation.} \label{mm2-tau}
 \end{figure}

The MM$^2$ distribution from Monte Carlo simulations of $e^+e^-\to
D^+D^-$, where the $D^-$ is fully reconstructed and
$D^+\to\tau^+\nu$, $\tau^+\to\pi^+\overline{\nu}$, is shown in
Fig.~\ref{mm2-tau}. While the pion in this decay sequence doesn't
have a narrow MM$^2$ peak as in the case of $D^+\to \mu^+\nu$,
many events are in the low MM$^2$ region. The spectrum peaks at
low MM$^2$ because the small $D^+$-$\tau^+$ mass difference causes
the $\tau^+$ to be almost at rest in the laboratory frame and thus
the $\pi^+$ has relatively large momentum. We must also ensure
that we do not accept $D^+\to\mu^+\nu$ events or semileptonic
decays with electrons.

Using our $D^-$ event candidates, we search for events with a single
additional charged track. The crystal calorimeter provides a way of
distinguishing this track among muons, pions and electrons. We
consider three separate cases: (i) the track deposits $<$~300 MeV in
the calorimeter, characteristic of a non-interacting pion or a muon;
(ii) the track deposits $>$~300 MeV in the calorimeter,
characteristic of an interacting pion; (iii) the track satisfies our
electron selection criteria defined below. Then we separately study
the MM$^2$ distributions for these three cases.

We exclude events with more than one additional, opposite-sign
charged track in addition to the tagged $D$, or with extra neutral
energy. Specifically, we veto events with extra charged tracks
arising from the event vertex or having a maximum neutral energy
cluster, consistent with being a photon, of more than 250 MeV.
These cuts are highly effective in reducing backgrounds especially
from $D^+\to \pi^+\pi^0$ decays.

The track candidates are required to be within the barrel region of
the detector $|\cos\theta|<0.81$. For cases (i) and (ii) we insist
that the track not be identified as a kaon.  For electron
identification we require a match between the momentum measurement
in the tracking system and the energy deposited in the CsI
calorimeter and the shape of the energy distribution among the
crystals is consistent with that expected for an electromagnetic
shower.

As demonstrated previously \cite{munuPRL}, the MM$^2$ distribution
has a shape well described by two Gaussians for the $\mu^+\nu$
mode with a resolution from Monte Carlo simulation (MC) of
0.0235$\pm$0.0004 GeV$^2$. We use different MM$^2$ regions for
cases (i) and (ii) defined above. For case (i) we define the
signal region to be the interval 0.175$>$MM$^2 >$0.05
 GeV$^2$, while for case (ii) we define the signal region to be the interval
 0.175$>$MM$^2 >$-0.05 GeV$^2$. Case (i) includes 98\% of the $\mu^+\nu$ signal,
 so we must exclude the region close to zero MM$^2$, while for case (ii) we are specifically
 selecting pions so the signal region can be larger.
 The upper limit on MM$^2$ is
 chosen to avoid background from the tail of the $\overline{K}^0\pi^+$ peak.
 The fractions of the MM$^2$
 range accepted are 46\% and 74\% for case (i) and (ii), respectively.

\begin{figure}[htbp]
 \centerline{
 \epsfxsize=4.0in \epsffile{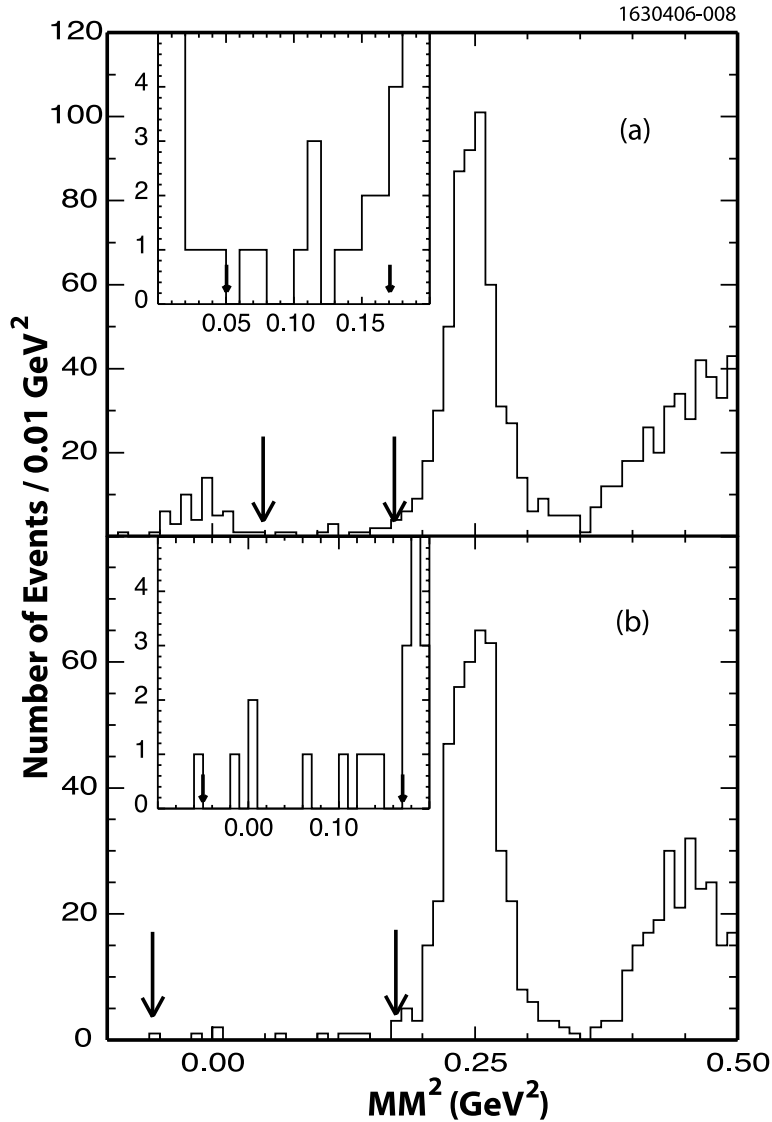}}
 \caption{The MM$^2$ distributions from data using $D^-$ tags and
 one additional opposite-sign
charged track and no extra energetic showers (see text). For the
case when the single track (a) deposits $<$~300 MeV of energy in
the calorimeter, case (i). The peak near zero is from
$D^+\to\mu^+\nu$ events. (b) Track deposits $>$~300 MeV in crystal
calorimeter but is not consistent with being an electron, case
(ii). The arrows indicate the signal regions. The insets show the
signal regions with a finer binning of 0.002 GeV$^2$.}
\label{mm2-tau-data}
 \end{figure}

\begin{figure}[htb]
 \centerline{
 \epsfxsize=4.0in \epsffile{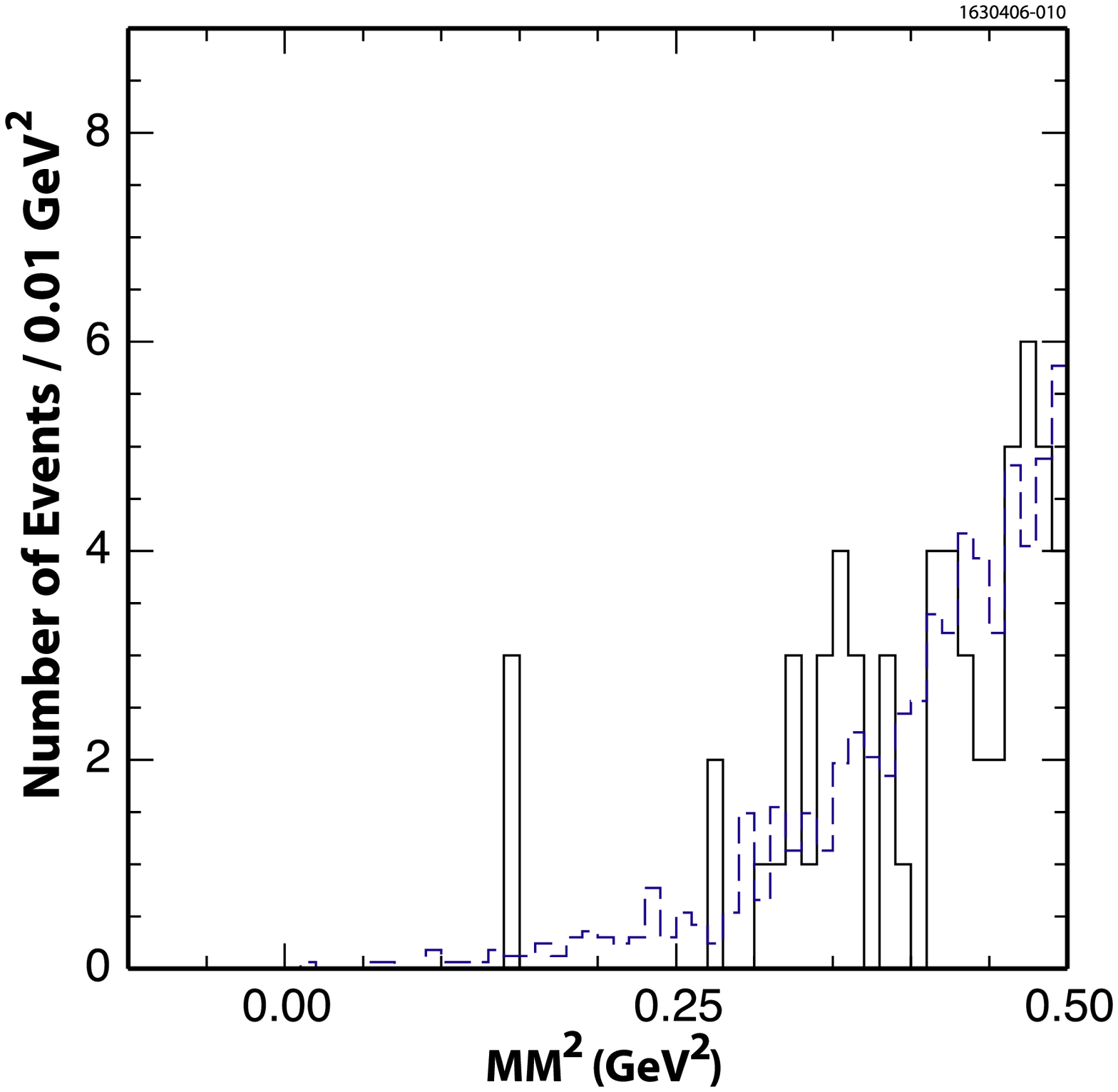}}
 \caption{The MM$^2$ distribution obtained using $D^-$ tags and one additional
opposite-sign charged track and no extra energetic showers (see
text). The track is required to deposit $>$~300 MeV of energy in
the calorimeter and be consistent with an electron. The solid
curve is data and the dashed curve Monte Carlo. }
\label{mm2-tau-data-elec-mc}
 \end{figure}

The MM$^2$ distributions for cases (i) and (ii) are shown in
Figs.~\ref{mm2-tau-data} (a) and (b). There are 12 events in the
signal region for case (i) and 8 for case (ii). The electron
sample, case (iii), shown in Fig.~\ref{mm2-tau-data-elec-mc}, has
3 events in the signal region and is used for background studies.

\section{Background Evaluation}
\subsection{Monte Carlo Estimates}
There are several background sources we need to evaluate. These
include background from other $D^+$ modes, background from
misidentified $D^0\overline{D}^0$ events and continuum background
including that from $e^+e^-\to\gamma\psi(2S)$, termed ``radiative
return."  There are a few $D^+$ decay modes that have been
identified {\it{a priori}} as possible background sources. These
are listed in Table~\ref{tab:Dpback}, along with the numerical
background estimates we obtain using Monte Carlo generation and
reconstruction of each specific mode. The branching ratios are
from the Particle Data Group except for the $\pi^+\pi^0$ and
$\rho^+\pi^0$ modes where we use new CLEO measurements
\cite{Blusk}. We note that often at least one photon from the
$\pi^0$ decay in these two modes exceeds our 250 MeV calorimeter
energy requirement and causes these decays to be vetoed.

\begin{table}[htb]
\begin{center}
\begin{tabular}{lccc}
    Mode & ${\cal{B}}$ (\%) & \# of events case(i)&\# of events case(ii)\\\hline
$\pi^+\pi^0 $ &0.12$\pm$0.01   & 0.13$\pm$0.02$\pm$0.01 & 1.40$\pm$0.07$\pm$0.11\\
$\overline{K}^0\pi^+ $&  2.77$\pm$0.18  & 2.44$\pm$0.51$\pm$0.17& 1.59$\pm$0.41$\pm$0.11\\
$\mu^+\nu $ &0.04$\pm$0.01   & 1.25$\pm$0.03$\pm$0.19&0.46$\pm$0.07$\pm$0.07\\
$\rho^+\pi^0$ &0.38$\pm$0.03   & 0.18$\pm$0.05$\pm$0.01 &0.23$\pm$0.05$\pm$0.02\\
$\pi^0 \mu^+\nu $ &0.44$\pm$0.07& 0.98$\pm$0.14$\pm$0.15 & 0.002$\pm$0.001$\pm$0.001\\
$\tau^+\nu$, $\tau^+\to\rho^+\nu$ &0.030$\pm$0.005   & 0.14$\pm$0.01$\pm$0.02 & 0.15$\pm$0.01$\pm$0.02\\
$\tau^+\nu$, $\tau^+\to\mu^+\nu \overline{\nu} $ &0.020$\pm$0.003   & 0.27$\pm$0.01$\pm$0.04 & 0.03(32\% C.L.)\\
Other $D^+$ modes & -&0.08(32\% C.L.) & 0.08(32\% C.L.)\\
$D^0$ modes & -&0.23$\pm$0.12$\pm$0.01 &0.42$\pm$0.16$\pm$0.01\\
Continuum & -&0.45$\pm$0.26$\pm$0.03 &0.74$\pm$0.33$\pm$0.05
\\\hline
Sum &  -&6.07$\pm$0.60$\pm$0.31&4.99$\pm$0.56$\pm$0.19\\
\hline\hline
\end{tabular}
\end{center}
\caption{Monte Carlo estimated backgrounds from all sources. The
second errors are systematic and are due to uncertainties on the
measured branching ratios for $D^+$ background sources and
production cross-section uncertainties for $D^0$ and continuum
sources. The ``other $D^+$ modes listed at 0.08 at 32\% c.l.
represent a 1$\sigma$ upper limit on this contribution.}
\label{tab:Dpback}
\end{table}

\begin{figure}[htb]
\centerline{ \epsfxsize=3.0in \epsffile{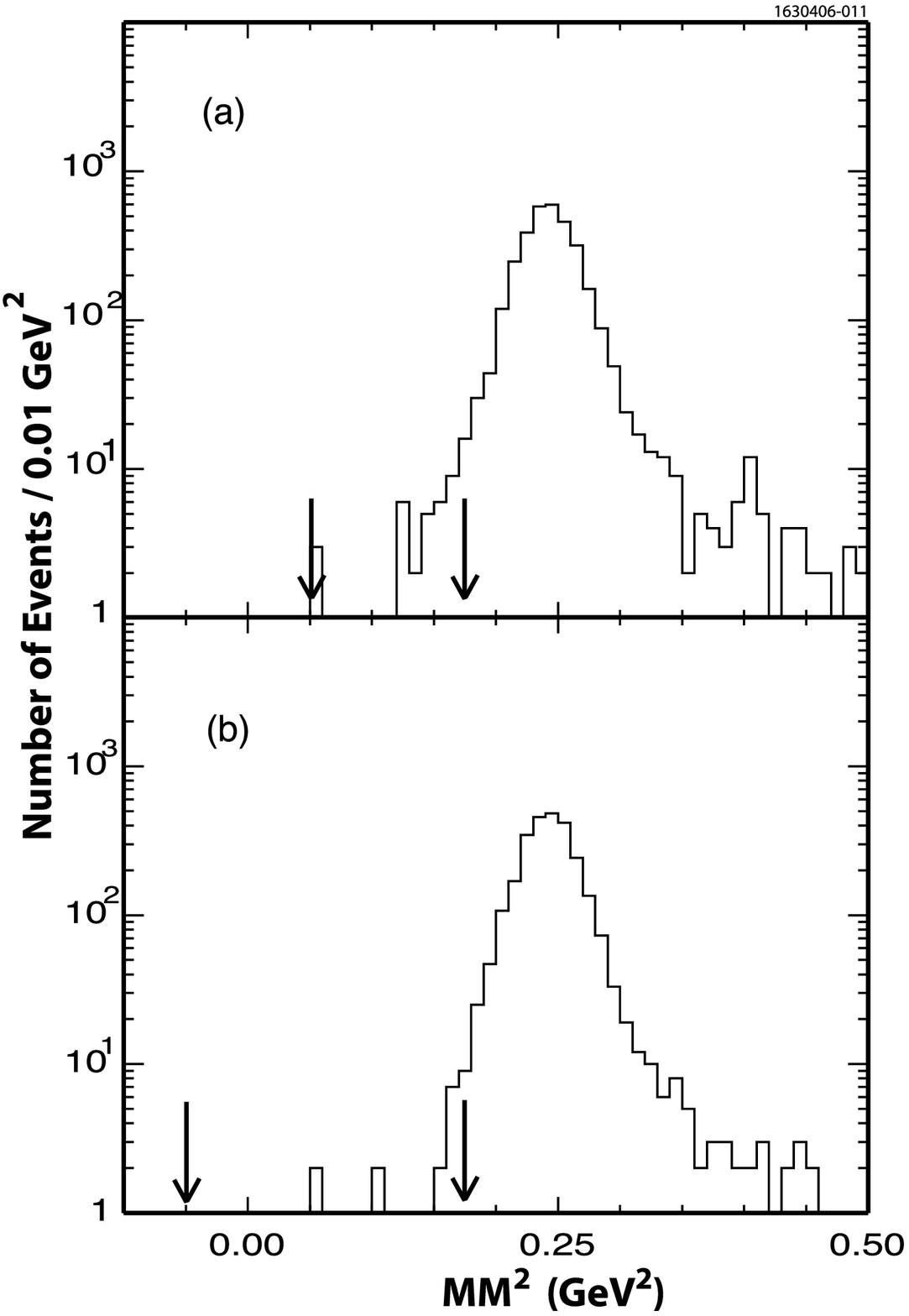} }
\caption{The MM$^2$ distribution from data events with a single
$D^0$ or $\overline{D}^0$ tag and the other neutral $D$ decaying
into two tracks, most likely $K^{\mp}\pi^{\pm}$, where the kaon
information is ignored. For the two cases: (a) track deposits
$<$~300 MeV of energy in the crystal calorimeter and (b) track
deposits $>$~300 MeV in the calorimeter. The arrows delineate the
relevant signal regions.}\label{K0tail}
\end{figure}

The $\overline{K}^0\pi^+$ mode gives a large peak in the MM$^2$
spectrum near 0.25 GeV$^2$. We need to evaluate the effects of the
tail of the distribution leaking into our signal region. A
simulation of this background for case (i) and case (ii) yields
$2.4 \pm 0.5 \pm 0.2$, and $ 1.6\pm 0.4 \pm 0.1$ events,
respectively. The systematic errors are due to uncertainties on
the measured branching ratios.

We have also checked the possibility of other $D^+D^-$ decay modes
producing background with an equivalent 1.7 fb$^{-1}$ Monte Carlo
sample; we find no additional events. {$D^0\overline{D}^0$} and
continuum backgrounds are evaluated by analyzing Monte Carlo
samples corresponding to 4.7 fb$^{-1}$ and 1.7 fb$^{-1}$,
respectively. To normalize our Monte Carlo events to our data
sample, we used $\sigma_{D^0\overline{D}^0}=3.6\pm0.1$ nb and
$\sigma_{\mathrm{continuum}}=14.5\pm1.0$ nb \cite{sighad}. Our
total background is 6.1$\pm$0.6$\pm$0.3 events in case (i) and
5.0$\pm$0.6$\pm$0.2 events in case (ii).

\subsection{Background Estimates From Data}
The largest source of background is the tail of the
$\overline{K}^0\pi^+$ peak. Simulations of the tails of
distributions, however, are often unreliable. Therefore, we also
measure this background rate directly from data.

We select $D^0\overline{D}^0$ events where one neutral $D$ decays
into $K^{\mp}\pi^{\pm}\pi^+\pi^-$, $K^{\mp}\pi^{\pm}\pi^0$ or
$K^{\mp}\pi^{\pm}$. These single-tag candidates are reconstructed
using tight selection criteria on $\Delta E$ and $m_{\mathrm BC}$.
In this sample, we look for events with only two additional
oppositely-signed tracks where the RICH system identifies one as a
kaon and the other as a pion. We insist that the charge of the kaon
candidate be opposite to the charge of the kaon in the tag mode. Our
aim is to isolate the $K^{\mp}\pi^{\pm}$ final state opposite the
reconstructed tag signal events. We avoid, however, making tight
cuts that might ameliorate the effects of tails.

The $K^{\mp}\pi^{\pm}$ final state is identical kinematically to
the $\overline{K}^0\pi^+$ state that we wish to emulate if we
ignore the measurements of the charged kaon and then compute the
MM$^2$, as shown in Fig.~\ref{K0tail} for cases (i) and (ii).

The event numbers in our signal ranges are $4.8 \pm 1.0 \pm 0.1$
for case (i) and $2.5 \pm 0.8 \pm 0.1$ for case (ii). The
systematic error arises from the normalization, derived from the
fit to the MM$^2$ peak near 0.25 GeV$^2$. There are backgrounds,
however, in these distributions from $\overline{D}^0\to\pi^+\pi^-$
and $\overline{D}^0\to\mu^-\pi^+\bar{\nu}$ events where the
candidate kaon is a misidentified pion. The probability for pions
faking kaons in this momentum range has been measured as
(1.10$\pm$0.37)\% \cite{fakes}. Using the known branching ratios
for the above two modes, we estimate 0.08 and 0.17 $\pi^+\pi^-$
events, respectively, and 0.01 and zero $\pi^+\mu^-\bar{\nu}$
events, respectively, that need to be removed from the background
estimate, leaving $4.7\pm 1.0$ and $2.4\pm 0.7$ background events.
This estimate is in reasonable agreement with the simulation.
(Since we are going to quote an upper limit in this paper,
choosing the Monte Carlo background estimate provides a worse
limit because less background is subtracted, and thus is the more
conservative choice.)

Another background check is to both measure the electron
background and simulate it. We note that the background due to
real muons should be almost equal to the background due to real
electrons. For this study we use the entire MM$^2$ region up to
0.5 GeV$^2$. The MM$^2$ distribution due to electron candidates in
the data is compared with the one from the Monte Carlo in
Fig.~\ref{mm2-tau-data-elec-mc}. There are 60$\pm$8 electrons in
the data compared with 63$\pm$3 in the Monte Carlo after
normalizing to the luminosity in the data. (In the signal region
there are 3 events in the data versus 3.9$\pm$0.1 in the Monte
Carlo.) The good agreement establishes that the Monte Carlo
properly predicts the semileptonic decay backgrounds.

\section{Branching Ratio Limits}

We do not observe a statistically significant difference between the
number of  signal and background events. For case (i) we have a net
signal of 5.9 $\tau^+\nu$, $\tau^+\to\pi^+\overline{\nu}$ signal
events and for case (ii) our yield is 3.0 events For each of our two
cases denoted by $j$, where $j$ represents either case (i) or case
(ii),  the expected number of events, $N_{\rm expected}^j$, is
related to the true
${\cal{B}}_{\tau\nu}\equiv{\cal{B}}(D^+\to\tau^+\nu)$ through the
relationship
\begin{equation}
N_{\rm expected}^j=N_{\rm tags} \times {\cal{B}}_{\tau\nu} \times
{\cal{B}}(\tau^+\to\pi^+\overline{\nu}) \times \varepsilon^j +
N_{\rm bkg}^j, \label{eq:N1}
\end{equation}
where $N_{\rm tags}$ is the number of single tag events and equals
160,729, after correcting for the slight difference in
reconstruction efficiency for tags opposite a single track versus
tags opposite a typical $D^+$ decay; $\varepsilon^j$ is the
efficiency, and $N_{\rm bkg}^j$ is the background. The Poisson
probability distribution $L_{j}(\cal{B}_{\tau\nu})$ for
${\cal{B}}(D^+\to\tau^+\nu)$ in each case $j$ has a mean equal to
$N_{\rm expected}^j$ and is given by:
\begin{equation}
{L_{j}({\cal{B}}_{\tau\nu})}=\left(1\over N^{j}!\right)\times
\exp\left(-N_{\rm expected}^{j}\right)\times \left(N_{\rm
expected}^{j}\right)^{N^{j}},
 \label{eq:L1}
\end{equation}
where $N^j$ is the number of detected $\tau^+\nu$ candidates:  8
for case (i) and 12 for case (ii).

Our results for the branching fractions are found by doing a
simultaneous likelihood fit of the distributions described in
Eq.~\ref{eq:L1}. We take into account the the different efficiencies
in cases (i) and (ii) that arises from both the MM$^2$ acceptance
(46\% and 74\%) and the efficiency of not having another unmatched
shower in the event with energy greater than 250 MeV (93.9\% and
91.8\%). We have previously found \cite{munuPRL} that the Monte
Carlo matched within 1.8\% our measurement of the extra unmatched
shower cut and thus use a slightly larger 2\% for the systematic
error on this quantity. Overall, the efficiencies are 18.7\%
(22.4\%), for case (i) and (case(ii)), respectively.

We find ${\cal{B}}(D^+\to\tau^+\nu)=(1.8^{+1.2 }_{-0.9} \pm 0.1
)\times 10^{-3}$ and $(0.8^{+0.9 }_{-0.5} \pm 0.2 )\times
10^{-3}$, for cases (i) and (ii), respectively, where the
statistical errors result from the values of ${\cal{B}}_{\tau\nu}$
corresponding to $34\%$ of the area under the $L_J$ distribution
above and below the maximum value.

 The errors on the backgrounds are treated as systematic and are obtained by
varying the background contributions in the likelihood distribution.
The systematic errors on the branching ratio from sources other than
backgrounds are listed in Table~\ref{tab:eff}; they are negligible
in comparison with the statistical uncertainty. A more detailed
explanation of the sources of systematic errors can be found in our
previous Letter \cite{munuPRL}.
\begin{table}[htb]
\begin{center}
\begin{tabular}{lc}
     &Systematic errors (\%) \\ \hline
MC statistics &0.2  \\
Track finding &0.7 \\
PID cut &1.0 \\
Minimum ionization cut &1.0 \\
Number of tags& 0.5\\
Extra showers cut & 2.0  \\\hline
Total &2.6\\
 \hline\hline
\end{tabular}
\end{center}
\caption{Systematic errors on the $D^+ \to \tau^+ \nu_{\tau}$
branching ratio.} \label{tab:eff}
\end{table}

To obtain a combined result for the branching fraction, we
construct the global likelihood as the product of the two Poisson
probability distributions, and we extract the value of
${\cal{B}}_{\tau\nu}$ which maximizes this likelihood function. We
find ${\cal{B}}(D^+\to\tau^+\nu)=(1.2^{+0.7 }_{-0.6} \pm 0.1
)\times 10^{-3}$. We caution the reader that this is not a
definitive measurement but an intermediate step used in the
process of forming an upper limit. (Had we used the data to
estimate the background, the branching fraction would be lower.)

Since the result is not statistically significant we quote an
upper limit of
\begin{equation}
{\cal{B}}(D^+\to\tau^+\nu) <2.1\times 10^{-3}~
\end{equation}
 at 90\%
confidence level.

The ratio to the expected rate in the Standard Model using our
measured ${\cal{B}}(D^+\to\mu^+\nu)$ is $< 1.8$ at 90\% confidence
level.

\section{Conclusions}
We have measured the first upper limit on the decay
$D^+\to\tau^+\nu$. We limit ${\cal{B}}(D^+\to\tau^+\nu)$ branching
ratio to $<2.1\times 10^{-3}$ at 90\% confidence level. We use our
previously measured result of
${\cal{B}}(D^+\to\mu^+\nu_{\mu})=(4.40\pm
0.66^{+0.09}_{-0.12})\times 10^{-4},$ \cite{munuPRL}, coupled with
the evaluation of Eq.~(\ref{eq:rat}) of 2.65, to limit the ratio
$R=\Gamma\left(D^+\to\tau^+\nu\right)/\Gamma\left(
D^+\to\mu^+\nu\right)$ to that expected in the Standard Model to
 $< 1.8$ at 90\% confidence level. Thus lepton
universality in purely leptonic $D^+$ decays is satisfied at the
level of current experimental accuracy.

\section{Acknowledgments}

We gratefully acknowledge the effort of the CESR staff in
providing us with excellent luminosity and running conditions. We
thank K. Agashe for useful discussions. D.~Cronin-Hennessy and
A.~Ryd thank the A.P.~Sloan Foundation. This work was supported by
the National Science Foundation, the U.S. Department of Energy,
and the Natural Sciences and Engineering Research Council of
Canada.

\end{document}